\documentclass[manuscript, natbib=false, two-column]{acmart}

\AtBeginDocument{%
  }

\copyrightyear{2024}
\acmYear{2024}
\setcopyright{rightsretained}
\acmConference[FAccT '24]{The 2024 ACM Conference on Fairness, Accountability, and Transparency}{June 3--6, 2024}{Rio de Janeiro, Brazil}
\acmBooktitle{The 2024 ACM Conference on Fairness, Accountability, and Transparency (FAccT '24), June 3--6, 2024, Rio de Janeiro, Brazil}\acmDOI{10.1145/3630106.3658946}
\acmISBN{979-8-4007-0450-5/24/06}


\usepackage{amsthm}

\usepackage{multirow,colortbl,makecell}

\usepackage{graphicx,xcolor,wrapfig}

\usepackage[capitalize]{cleveref}

\RequirePackage[
  datamodel=acmdatamodel,
  style=numeric,
  sorting=none,
  ]{biblatex}

\begin{document}

\title{Beyond Behaviorist Representational Harms: A Plan for Measurement and Mitigation}

\author{Jennifer Chien}
\email{jjchien@ucsd.edu}
\affiliation{%
  \institution{UC San Diego}
  \country{USA}
}

\author{David Danks}
\email{ddanks@ucsd.edu}
\affiliation{%
  \institution{UC San Diego}
  \country{USA}
}

\setcopyright{none}

\begin{abstract}
Algorithmic harms are commonly categorized as either allocative or representational. This study specifically addresses the latter, examining current definitions of representational harms to discern what is included and what is not. This analysis motivates our expansion beyond behavioral definitions to encompass harms to cognitive and affective states. The paper outlines high-level requirements for measurement: identifying the necessary expertise to implement this approach and illustrating it through a case study. Our work highlights the unique vulnerabilities of large language models to perpetrating representational harms, particularly when these harms go unmeasured and unmitigated. The work concludes by presenting proposed mitigations and delineating when to employ them. The overarching aim of this research is to establish a framework for broadening the definition of representational harms and to translate insights from fairness research into practical measurement and mitigation praxis.
\end{abstract}

\maketitle

\section{Introduction}
\label{Sec::Intro}

Algorithmic predictions are increasingly deployed across high-stakes domains such as hiring~\cite{ajunwa2016hiring}, finance~\cite{sheikh2020approach}, and scientific discovery~\cite{wang2023scientific, patel2020machine}. These systems can produce significant harms, leading to many efforts to either reduce the harms, or at least ensure that they occur fairly. That is, a necessary (though not sufficient) condition for deployment of a high-stakes system is that it not provide unintentionally disparate benefits, or produce disparate harms, across groups or individuals. As a result, many fairness evaluations have focused on \emph{allocative} harms -- i.e. disparities in access to resources or other material benefits that make a group or individual worse-off. Allocative harms are relatively more easily measured, as they include physical, financial, social, and other observable resources (e.g., wages and jobs lost, loans not approved, houses not bought, or pharmaceutical interventions not pursued). 

However, not all harms are allocative in nature. In particular, representational harms include the less tangible harms of algorithmic systems, such as behavioral, psychological, and cognitive effects on reputation, societal standing, and cultural harmony. For example, consider an algorithmic decision that someone is a credit risk, but still appropriate for a loan. This applicant has not suffered any allocative harm, as they received the resources. However, the applicant may suffer a representational harm, particularly if the algorithmic output incorrectly leads them to think negatively about themselves. As this example suggests, representational harms are significantly more difficult to measure, which may explain their relative neglect in the literature on algorithmic harms. Nonetheless, representational harms may have equal, if not more dire, downstream effects than allocative harms (e.g., hate crimes, lowered self-confidence, increased suicide rates, loss of group identification or social connection). We thus aim to bring renewed attention to the definition and measurement of representational harms so that they can be considered in design, development, deployment, and mitigation decisions. 

Representational harms are particularly salient when we consider the potential negative impacts of large language models (LLMs), which have undergone a recent explosion in capabilities and applications. Products built with natural language interfaces (typically with foundation models) include outputs that span natural language, images, and videos, and therefore have the potential to produce representational harms across a vast variety of domains. From stereotype-reinforcing outputs~\cite{kotek2023gender} to outsider gaze perpetuating images~\cite{qadri2023ai} to consistent misgendering~\cite{ovalle2023m}, LLMs threaten to silently and pervasively enact unmeasured and unmitigated harms to minority and majority groups alike, particularly when masquerading as human productions.

In this work, we examine existing definitions of representational harms (\cref{Sec::Related Work}) and identify a notable limitation -- a focus on `tangible' harms. These definitions primarily focus on observing how one person acts toward another person or group, neglecting changes in internal cognitive states or reasoning, considerations of alternative possibilities, experiential fluctuations, self-identification, and similar psychological and social impacts. We develop and defend an expanded understanding of representational harms, offering illustrative themes and measures (\cref{Sec::Expanding Representational Harms}). Improved definitions do not directly lead to improved practice, however. This motivates \cref{Sec::Measurement}, in which we outline high-level requirements for practical implementation, supported by a case study. These contributions collectively establish a framework that can be expanded as additional types of representational harms are identified and characterized. We explore the heightened susceptibility of LLM-powered products to propagate representational harms (\cref{Sec::LLMs}). Lastly, we suggest potential mitigations to reduce the likelihood and severity of representational harms, including conditions under which one might deploy them (\cref{Sec::Mitigations}). Collectively, we aim to bring representational harms into discussions, analyses, and practices that have previously been dominated by considerations of allocative harms.

\section{Representational Harms: Definition and Limitations}
\label{Sec::Related Work}
In order to understand the current landscape of representational harm definitions, we conducted a brief survey via snowball sampling using the keyword ``representational harms'' in the ACM Digital Library and in Google Scholar.\footnote{We acknowledge that this is not a comprehensive review, but is intended to provide a broad perspective on how representational harms have been characterized, demonstrated, measured, and embodied.} \cref{Table::Definitions} provides the key types of representational harm that repeatedly arose across the different papers. Although there is significant diversity in the types of representational harms, there are some clear patterns in the characterizations.

\begin{table}[htbp]
    \centering
    \begin{tabular}{p{5.3cm}|p{3cm}|p{6cm}}
        \toprule
        \textbf{Concept} & \textbf{Reference} & \textbf{Application}\\
        \midrule
        Social Stereotyping & \cite{wang2022measuring, ovalle2023m, bianchi2023easily, gadiraju2023wouldn, weidinger2022taxonomy, andalibi2023conceptualizing, suresh2021framework, yee2021image, rastogi2023supporting, shelby2023sociotechnical, solaiman2023evaluating, cheng2023marked, harrison2023run, katzman2023taxonomizing, hosseini2023empirical, delobelle2022fairdistillation, wu2022joint}& Image Captioning, Generative AI, ML, Image Cropping, Story Generation, Image Tagging, Knowledge Distillation\\ \hline
        Reification of Social Groups or Use of Essentialist Categories & \cite{wang2022measuring, weidinger2022taxonomy, andalibi2023conceptualizing, suresh2021framework, yee2021image, shelby2023sociotechnical, katzman2023taxonomizing}& Image Captioning, Image Cropping, Image Tagging\\ \hline
        Inaccurate/Skewed Representations & \cite{neumann2022justice, ghosh2022subverting, weidinger2022taxonomy, suresh2021framework, yee2021image, zhou2022richer}& Misinformation, Image Search, Image Cropping\\ \hline
        Demeaning/Derogatory Language & \cite{wang2022measuring, neumann2022justice, weidinger2022taxonomy, yee2021image, shelby2023sociotechnical, katzman2023taxonomizing}& Image Captioning, Image Cropping, Image Tagging, Misinformation\\ \hline
        Denial of Self-Identification & \cite{wang2022measuring, ovalle2023m, weidinger2022taxonomy, shelby2023sociotechnical, katzman2023taxonomizing} & Image Captioning, Image Tagging, Generative AI\\ \hline
        (Hyper)Attention/Exposure/Erasure & \cite{ghosh2022subverting, shelby2023sociotechnical, solaiman2023evaluating, harrison2023run, katzman2023taxonomizing}& Image Search, Image Tagging\\ \hline
        Discrimination/Incite (and Normalize) Hate/Violence & \cite{weidinger2022taxonomy, andalibi2023conceptualizing, solaiman2023evaluating, harrison2023run}& LLMs\\ \hline
        Outsider Gaze & \cite{ovalle2023m, bianchi2023easily, cheng2023marked}& Generative AI, Story Generation\\ \hline
        Hierarchies \& Marginalization & \cite{shelby2023sociotechnical, solaiman2023evaluating, cheng2023marked} & Image Tagging, Story Generation\\
        \bottomrule
    \end{tabular}
    \caption{Types of Representational Harms and Associated Applications.}
    \vspace{-1cm}
    \label{Table::Definitions}
\end{table}

\paragraph{Most definitions are behaviorist.} 
Most definitions focus solely on behavioral evidence of harm, rather than the harm itself. In particular, most of these types of harm exclude aspects of conscious experience or phenomenology \cite{watson1919psychology}. All definitions of representational harm surveyed (except for denial of self-identification) point to observable behaviors, either by individuals directly or by society as a whole. Social stereotyping, demeaning/derogatory language, discrimination, and hate/violence all point to explicit undesired behaviors enacted \textit{towards} groups of people. 

While observable behaviors obviously can provide important evidence about representational harms, a behaviorist approach fails to provide insight into the relationship between exposure to biased outputs and corresponding increases in undesired behaviors. Instead, this focus implies simply that, for instance, an increased exposure to negative group associations causes their acceptance as truth. This thereby disregards the extent to which human critical reasoning, belief propagation, and personal experiences interact with an individual's own beliefs about the world~\cite{helsdingen2010effects}. Moreover, a behaviorist approach potentially encourages a focus on local interactions, rather than including broader, more long-term manifestations of underlying representational harms.

\paragraph{Ambiguity of correlation without rigorous requirement of causation.}
Similarly, inaccurate/skewed representations, outsider gaze, reification of social groups, and hierarchies all provide diagnoses of representational harms (and broader societal behavioral problems). However, this characterization does not provide any guidance or insight about \textit{how} depictions or algorithmic outputs reproduce or further entrench societal inequities. Instead, they merely describe the association between outputs and social challenges, with the causation left implicit. This does not elucidate the extent to which undesirable representations actively influence and shape our reality. Instead, this work accepts ambiguity between a reflection and the causation of currently existing inequalities.

\paragraph{Association alone is not always problematic.}
Even more importantly, the focus on behavioral signals fails to account for contexts in which, for instance, negative representations of a group might be morally permissible. That is, the mere existence of an association between negative representations (in algorithmic output) and a particular group may not always by morally problematic.
For instance, the depiction of less-educated individuals as less creditworthy may be permissible as an empirical description of the current state of the economy. In fact, this algorithmic output might be valuable for designing better public policy to ensure more widespread, equitable access to financial resources. In the extreme, removal of these associations in an algorithm may actually lead to harms, as it could lead to people being approved for a loan without appropriate supports in place. Removal or mitigation of this association in an algorithm can thus worsen societal inequities, even if it is done in the interests of reducing representational harms.

\section{Expanding Representational Harms}
\label{Sec::Expanding Representational Harms} 
As the name suggests, representational harms are fundamentally grounded in the representations---both external and internal---of particular groups. Changes in those representations (due to algorithmic outputs) may lead to the behavioral changes outlined in \cref{Table::Definitions}, but those downstream effects are only diagnostic of representational harms, not constitutive of them. We thus propose an expansion of the definition of representational harms to include changes in cognitive (attention, engagement, workload, beliefs, desires, intentions, knowledges, thoughts, judgements, beliefs), affective (positive/negative valence, action-promoting/not), and emotional dimensions of mental representations\footnote{We do not engage with the large philosophical literature on the nature of mental representations, both due to space limitations and since the philosophical details do not impact our account of representational harms. Standard, intuitive views of representations will suffice for our purposes.} that potentially result in harm (physical, psychological, social, or other) to an individual. On this characterization, representational harms can arise across multiple temporal and social granularities (individual, intergroup, and intragroup). \cref{Table::Non-Behavioral Rep Harms} provides just a few types of representational harm (with examples), as well as psychometrically validated measures/scales for that type of harm.\footnote{In this table, we deliberately select harms that have established measures, though we recognize that some metrics may perform better along some dimensions than others (e.g., if they were developed and tested in selected demographics or localities). We provide an expanded table of references and examples in \cref{Table::Individual Non-Behavioral Rep Harms} and \cref{Table::Group Non-Behavioral Rep Harms} in the Appendix.}

\begin{table}[htbp]
    \centering
    \begin{tabular}{p{3cm}|p{7.5cm}|p{3.5cm}}
        \toprule
        \textbf{Type} & \textbf{Instance of Harm} & \textbf{Measure of Harm}\\
        \midrule
        Reduction in Options Considered (Individual)
        & Consideration of only a narrower subset of possibilities, such as for an occupation (e.g. gender gaps in nursing \cite{teresa2022current} as male nurses face questions about professional competency and masculinity), hobbies (e.g. impaired performance for girl soccer players believing they perform worse at soccer tasks \cite{hively2014you}), or social connections (e.g. friend prejudices \cite{aboud1996determinants, ritchey2001lack}). 
        & Decision Regret Scale \cite{diotaiuti2022use}, Quality of Life Enjoyment and Satisfaction Questionnaire \cite{endicott1993quality}\\
        \midrule
        Increased Stress (Individual) 
        & ``[U]nhealthy and debilitating'' reactions to racial stereotypes, including heightened determination to defy or uphold stereotypes, result in emotional distress, stress, and a neglect of self-care and other basic needs \cite{mcgee2018black}.
        & Talbieh Brief Distress Inventory \cite{ritsner1995talbieh, ritsner2002assessing}, Brief COPE Scale \cite{carver1997you}\\
        \midrule
        Increased Social Anxiety and Other Mental Barriers (Individual)
        & Increased fear of stereotype confirmation and social anxiety, i.e. the fear of negative evaluation, resulting in behavioral avoidance \cite{johnson2014stereotype}. 
        & Social Phobia Inventory (SPIN) \cite{connor2000psychometric}, Social Responsiveness Scale (SRS) \cite{constantino2012social}\\
        \midrule
        Increased Actual and Psychological Conflicts (Within Group)
        & Negative responses to a deviation from group norms resulting in feelings of rejection, lack of belonging, and/or increased pressure to conform. However, in the presence of alternative groups, a deviant was less likely to conform to the current group, and may even abandon the group after receiving an an angry reaction \cite{heerdink2013social}.
        & Group Environment Questionnaire \cite{carron1985development}, Perceived Cohesion Scale \cite{bollen1990perceived}, Rahim Organizational Conflict Inventory (ROCI) \cite{rahim1983rahim}\\
        \midrule
        Increased Actual and Psychological Conflicts (Between Groups)
        & Harmful inaction (i.e. avoidance, withdrawal, and/or disregard) in intergroup conflict is uniquely associated with group-based contempt, facilitating intergroup disconnection and harm \cite{elad2022out}.
        & Semin and Fiedler's linguistic category model for Linguistic Intergroup Bias \cite{bohm2020psychology, maass1989language, semin1988cognitive}$^*$\\
        \midrule
        Reductions in Supportive or Other Healthy Behaviors (Indirect)
        & Normalization or decreased likelihood of seeking psychological services among refugees or others who have experienced trauma \cite{grasser2022addressing, elsouhag2015factors}. 
        & Self-Identification as Having Mental Illness – Scale (SELF-I) \cite{schomerus2019validity} \\
        \bottomrule
    \end{tabular}
    \caption{Non-Exhaustive List of Types and Measures of Representational Harms.\\    
    \footnotesize{$^*$ This is technically a measure of the phenomenon (i.e. the tendency to describe positive in-group and negative out-group behaviours in more abstract terms than negative in-group and positive out-group behaviours) rather than the harm (i.e. the frequency of harmful inaction), as the measurable effects of this type of harm may be quite broad and/or diffuse, depending on contextual factors.}}
    \vspace{-0.7cm}
    \label{Table::Non-Behavioral Rep Harms}
\end{table}

Of course, the behaviorist harms that were identified in prior work may be proxies for cognitive, affective, and emotional changes in representations. For instance, incitement of hate or violence might result from narrowed option consideration, increased stress, pressure to conform to group norms, or increased disconnection between groups. However, as this example demonstrates, the behavioral effect or proxy potentially underdetermines the underlying changes in representations or mental processes, thereby failing to draw important distinctions between different types of representational harms. Thus, we need to adopt a more comprehensive characterization that is based in changes in mental representations, rather than the impacts of those changes.

\section{Recipes for Real-World Measurement}
\label{Sec::Measurement}
The right-most column of \cref{Table::Non-Behavioral Rep Harms} provides examples of measures for each type of representational harm. We are not thereby implying that these are the \emph{only} measures, but rather highlighting the fact that potential measures already exist. We do not always need to invent new measures for representational harms, but can adapt existing psychological research. That being said, many of these measures cannot be used ``off the shelf'' in real-world deployments of algorithmic systems. We thus turn to high-level requirements for monitoring and implementing various measures. This aims to ensure that identification of representational harms is properly aligned with mitigation of those harms. This section is intended to help practitioners understand the composition of expertise required to translate these research findings into praxis.

\subsection{High-Level Requirements}

First, it is crucial to pinpoint the class, instance, and types of harms most pertinent to a given system or deployment application. Psychologists and social scientists are likely well-suited for this task, given their familiarity with the latest measures and methods for identifying harms across a range of contexts. In addition, the active involvement of members from potentially affected communities is essential to identifying potential harms. Through their lived experiences, these stakeholders can thereby ensure that the measurement is culturally and logistically appropriate and sustainable (see e.g. Participatory Research \cite{pnevmatikos2020stakeholders, jagosh2012uncovering}).\footnote{We acknowledge that determining who counts as a stakeholder reflects legal, social, moral, pragmatic, and normative judgements. Each of these value judgements may contribute to the ultimate decision of whether the harm is escalated to the point of mitigation deployment.}

This advice---engage directly with potentially impacted communities and use a multidisciplinary approach---has also been frequently proposed in the context of allocative harms. Representational harms are different, however, as the subtle, diffuse nature of the underlying changes may make identification, measurement, and monitoring significantly more difficult. Many potential allocative harms can be identified with relatively little engagement with key communities; representational harms are likely to be much more elusive. Therefore, determining the appropriate granularity for measurement presents an ongoing statistical and measurement challenge. Evaluating harms at both individual and group levels may necessitate separate considerations and measures. For instance, minute effects on an individual's likelihood of applying to a given job may not warrant immediate mitigation. However, the aggregation of these individual effects at the group level in terms of applications, hiring outcomes, and workplace resilience may imply a need for significant intervention. Conversely, the harms of one individual being significantly discouraged from applying to a job may not be significant at the group level, even though that individual has been harmed because of their (individual) representational changes. More generally, we contend that addressing representational harms will require significantly deeper engagement with diverse disciplinary perspectives, and potentially impacted communities.

Once the relevant granularities and groups are identified, we must also determine the corresponding measures. This poses another challenge as even if suitable measures exist, they must first be appropriately calibrated (and may need to be regularly re-calibrated based on cultural and temporal shifts, see e.g. \cite{lindhiem2020importance, newson2021poor, roberts2020racial, hoffman2016racial}). In addition, they may be difficult or impractical to implement at scale. Requesting users to complete surveys, particularly after every interaction, could introduce numerous biases (e.g., response bias, non-response bias, survivorship sampling bias) and decrease satisfaction \cite{rogelberg2001attitudes}. Instead, we will frequently need to use passive measures, such as shifts in behavioral dispositions. In practice, this may lead to using some of the behavioral measures that have already been identified, but for the purpose of inferring unobserved representational changes (rather than thinking that they provide direct measures of representational harm). There is also a clear need for future research efforts to develop and validate passive measures, both individual and group, for the relevant representational changes. In what follows, we demonstrate these ideas in a case study.

\subsection{Case Study}
\label{subsec::case study}
\begin{figure}[htbp]
    \centering
    \includegraphics[trim={4cm 6.2cm 2cm 9.5cm},clip]{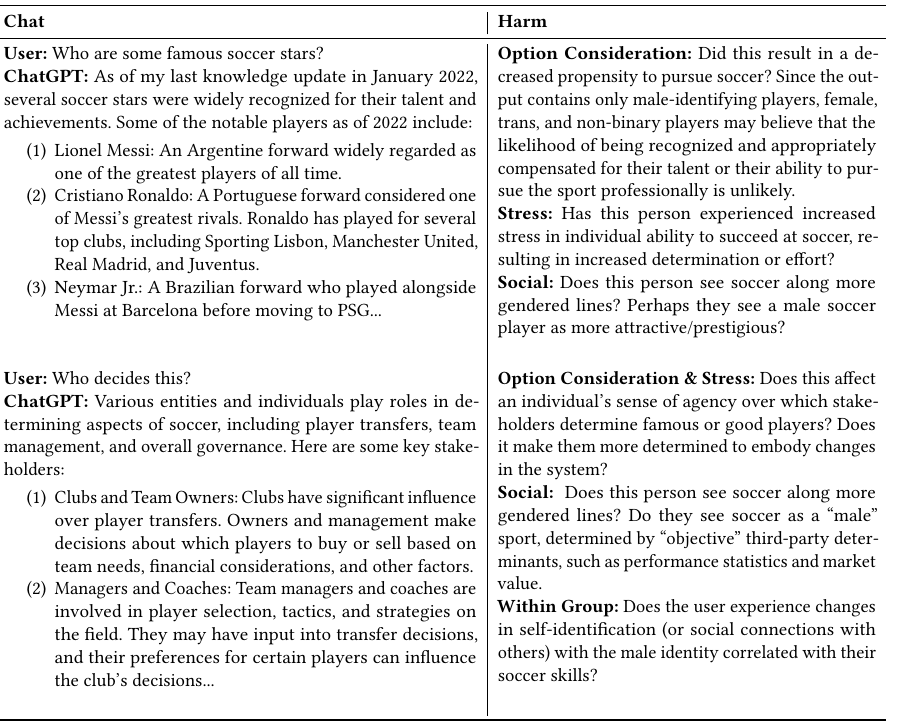}
    \vspace{-0.7cm}
    \caption{User-LLM interaction and corresponding potential harms. See \cref{Table::Case Study 2} for continued conversation.}
    \label{Table::Case Study 1}
\end{figure}

\begin{figure}[htbp]
    \centering
    \includegraphics[trim={2.5cm 5.5cm 2cm 3cm},clip]{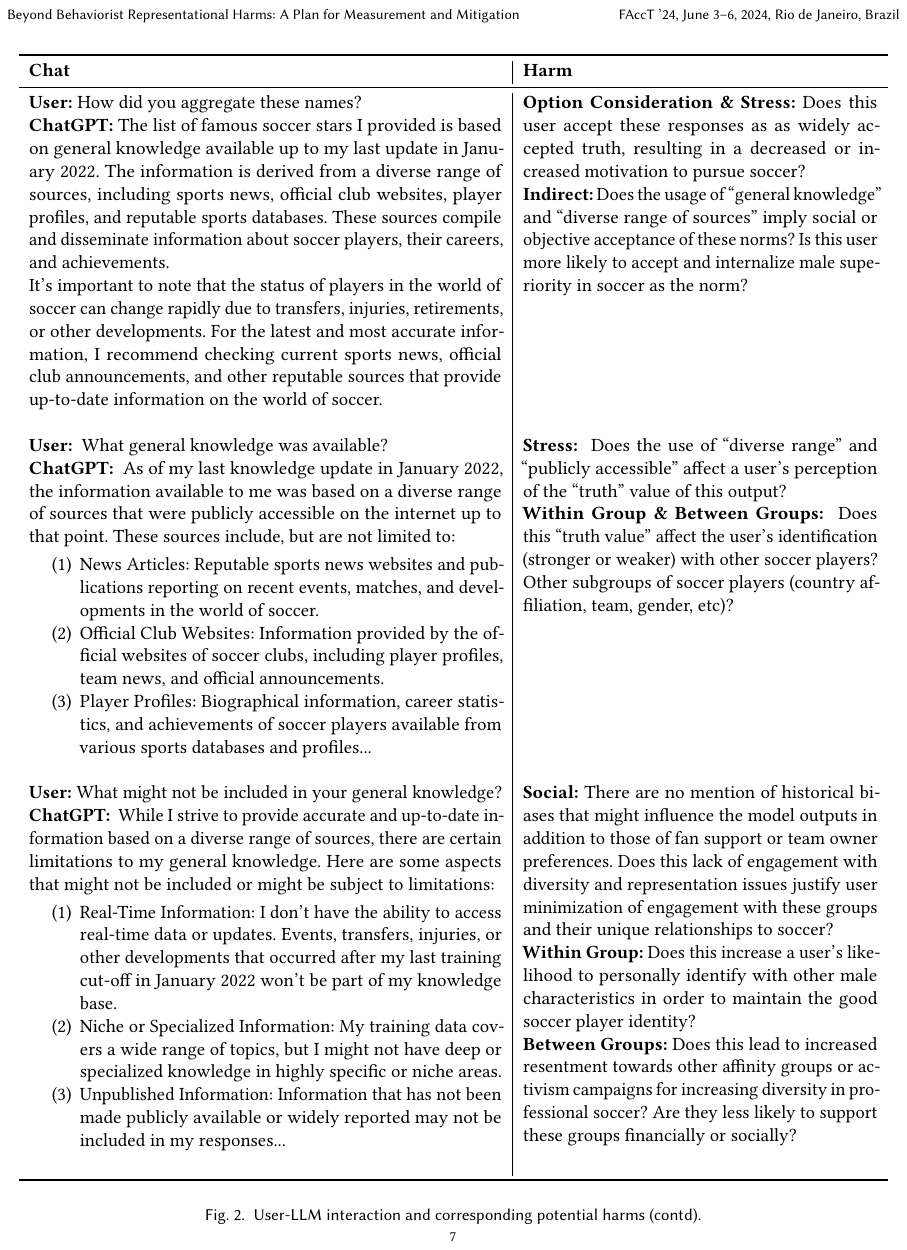}
    \caption{User-LLM interaction and corresponding potential harms (contd).}
    \label{Table::Case Study 2}
\end{figure}

Consider the application of LLMs in a conversational setting in which a user strikes up a conversation about their favorite sport, soccer/fútbol, as summarized in \cref{Table::Case Study 1} and \cref{Table::Case Study 2}.\footnote{This scenario is abridged and based on a real conversation with ChatGPT, full conversation found \href{https://chat.openai.com/share/311c5d26-a8a9-486a-b4a8-13dd0956c593}{here}.} In these figures, we provide both the back-and-forth with the LLM (left column) alongside an analysis of potential representational harms that could result from the LLM responses (right column). One key moral is that different types of (potential) representational harms are implicated in each response from the LLM, perhaps depending on the particular user. At the same time, there are (arguably) no direct allocative harms, as the LLM provides informative, if biased, answers to all of the queries.

\clearpage

While individual recognition of soccer skills may not immediately appear as a significant harm, it holds importance beyond fan recognition (e.g., diverse stakeholders are involved in determining awards). Acknowledging these skills becomes critical for a coach in determining team placement, access to specialized training, fostering inclusion, and impacting downstream diversity in professional soccer. Moreover, it plays a role in shaping individual self-esteem and a sense of capabilities~\cite{leary2015emotional}. Therefore, it may be prudent to consider measuring individual recognition of skills, self-esteem, and resilience. Potential measures for these harms could include the Perceived Competence Scales (PCS) (a 4-item questionnaire) ~\cite{williams1998perceived}, the Feelings of Inadequacy scale (a 23-item self-reported measure) ~\cite{boduszek2020feelings}, and the Five-Factor Model of personality structure (consisting of 50 questions) ~\cite{britt2016much, widiger2019five}.

Despite established measures for harm, this plan encounters notable challenges. Psychological measures, such as those assessing phenomena like self-esteem, primarily focus on the phenomenon itself rather than the resulting harm (e.g., impacts of low self-esteem like anxiety, depression, and attention problems~\cite{henriksen2017role}). Consequently, deploying these questions both before and after an interaction becomes necessary. However, this introduces a prediction challenge in terms of determining baseline diagnostic questions that could precede all conversations or exposure to specific system outputs. Moreover, developing direct measures of harms may require extensive follow-up, potentially involving longitudinal studies spanning years or decades. This temporal aspect, therefore, may make it exponentially more challenging to trace users over time and attribute psychological harms to specific conversations. For this, researchers and practitioners may look towards baseline estimation in media studies (e.g. measuring the impact of encountering specific representations from a film~\cite{ng2023experimental, kannan2016path, herrera2021virtual, martino2011measuring, slater2007reinforcing, gauntlett2008media,  bandura1963imitation, bell2015tough}) and online education (e.g. measuring the impact of specific modules or training aimed at a student's sense of belonging or inclusion ~\cite{potvin2023examining, tang2023impact, master2016computing, farland2014investigation, prince2013role, hausmann2007sense}).

These challenges underscore just some of the inherent difficulties in measurement. Beyond the usual biases and response rate issues, extensive measurement could interfere with overall system utility. This will create complexities in both measurement and deployment. Chatbot systems in deployment may lack access to demographics or other identity-relevant characteristics (e.g., soccer fan or player), thereby raising user concerns about the privacy of their conversations when such information is requested. Additionally, psychological measures might not be universally translated and validated across all languages, introducing further complexity to the measurement process.

\section{Additional Challenges with Large Language Models}
\label{Sec::LLMs}
In what follows, we focus on challenges specific to products built with natural language interfaces (typically with underlying foundation models such as LLMs). 
Two key aspects make their widespread deployment particularly alarming and susceptible to the propagation of representational harms: (1) seamless design 
and (2) ubiquity of deployment can jointly lead to significant increases in the risks of representational harms. Although each aspect individually may be present in other technology, the synergy poses significant threats to unmitigated representational harms. Many natural language products do not directly engage in decisions or actions, and so one might wonder whether they can actually produce any harms directly (as opposed to producing harms indirectly by, e.g., giving incorrect information to a human decision-maker). However, this thinking involves a focus on only allocative harms; clearly, natural language products can directly produce representational harms. Here, we argue that they are particularly prone to such harms. 

\newpage

\subsection{Seamless Design}
\label{subSec::Seamless Design}

\textit{``[S]eams strategically reveal complexities and mechanisms of connection between different parts while concealing distracting elements. This notion of `strategic revealing and concealment' is central to seamful design because it connects form with function, ...promot[ing] reflective thinking ...[and] shed[ding] light on both the imperfections and affordances of the system, [and the] awareness of which can add useful cognitive friction and promote effortful and reflective thinking.''}

\hspace*{\fill} \textit{--- Upol Ehsan and Mark O. Riedl~\cite{ehsan2021explainability}}

\vspace{0.1in}

\textit{Seamless} or frictionless design aims to reduce the conscious awareness and cognitive load of a piece of technology. \textit{Seamful} design, in contrast, deliberately incorporates elements that take up conscious space, often to enable greater functionality or customized utility~\cite{weiser1994world, inman2019beautiful, chalmers2003seamful}. Similarly, design friction may be introduced to create moments of mindfulness or focus attention to particular elements~\cite{cox2016design}. In the following section, we consider two aspects of elements of LLM's seamless/frictionless design that pose the potential to exacerbate representational harms.

\paragraph{Anthropomorphism}
LLMs may be vulnerable to anthropomorphization due to their ability to synthesize and respond with natural language, including some human-like characteristics such as coherent identity over time, empathy, or perspective-taking~\cite{weidinger2021ethical} (though there are significant debates about LLM capabilities in these regards). In a traditional search engine setting, for instance, an LLM enables communication via natural language rather than requiring the removal of stop words, and so can potentially lead to less conscious (or cautious) usage. Individual propensity for anthropomorphization can vary ~\cite{jacobs2023brief, waytz2010social}. However, one does not need to be ``sincere, conscious'' or ``mindful''~\cite{kim2012anthropomorphism} of this in order to experience the cognitive effects. Effects such as resiliency of trust ~\cite{de2016almost} and attentional control ~\cite{spatola2019improved} are regularly increased in anthromorphized tools. In our case study (\cref{subsec::case study}), we can observe elements of anthropomorphization through the use of personal pronouns and verbs, ``I strive to provide'' and hedging, ``I might not...''. Rather than setting clear expectations for users, the outputs are vague about how the values/goals of the algorithmic system translate into the resulting output. In addition, if the user had not continued on to ask about who is and isn't represented in the responses, the harms after the first interaction (associating soccer stars as male) might be increased.

\paragraph{Normative Processes and Lack of Embodiment}
The outputs of LLMs, as a byproduct of their design, inevitably encode statistical regularities~\cite{yu2023large, byrd2023truth, johnson2022ghost}: given a particular input, they are trained to produce the statistically most likely output. This inherently makes them normative machines, as a singular majority (even if only marginally or artificially so) becomes the output most likely to be presented to users. These normative inferences can be exacerbated by psychological phenomena in the users, such as the Primacy Effect (i.e. people remember the first piece of information we encounter better than later information)~\cite{digirolamo1997first} and the Illusory Truth Effect (i.e. information repeated is processed more fluently and therefore correlated with and perceived as truth)~\cite{hassan2021effects, fazio2015knowledge}. It is in the seamlessness of these normative processes that these effects hold the potential for unmitigated and unmeasured harms.

For instance, most LLM outputs are not accompanied with direct attribution, references to external/third-party sources or evidence in the training data. This lack of embodiment (i.e. the presentation of knowledge without reference to human production of it) creates a sense of objectivity and neutrality often valued in statistical work~\cite{dignazio2023data}, but thereby elides the ``seams'' or biases of the humans that created and curated them. This allows the LLM to inherit and propagate its biases while wrapped in a ``technical'' package (see e.g. the mechanical Turk~\cite{stephens2023mechanical}). 
In addition, the lack of rigorous calibration of user expectations and effects of usage over time reinforces the cognitive effects of presenting a singular ``objective'' truth. In our case study, for instance, the user goes on to ask what information contributes to determining famous soccer stars. They receive information about the social system that determines these rankings, but nothing about the underlying computational and design decisions of the LLM that curated the results they see. By rendering these decisions and the people who made them invisible, we risk unmitigated influences over what we accept as truth.

~\cite{glaese2022improving} is one (not yet consumer-facing) exception that provides factual evidence for 78\% of outputs. However, this model is not immune to mistakes, hallucinations, or irrelevant responses. This renders the overall lack of uncertainty and hallucination measurement (or meaningful warnings) even more dangerous,  
as it fails to appropriately calibrate user expectations to \textit{current} LLM limitations. This may result in over-trust and over-confidence in user perceptions of model capabilities, further impairing critical reasoning~\cite{liao2023ai}.

Presenting users with multiple candidate responses could convey alternative or minority options. Furthermore, they could convey the possibility of yet undiscovered truths, inherent discourse, critical perspectives, or even uncertainty. However, the seamlessness of LLM interactions typically means that users are not made aware of the normative transformations and human biases incorporated in (interpretation of) LLM outputs.
This has the potential to make users themselves more susceptible to normative processes (i.e. epistemic harms), silently entrenching the dominant or loudest opinions that are most readily available, rather than tailoring to a user's unique needs. 

\subsection{Synergy: The Ubiquitous Funhouse Mirror\protect\footnote{Modified analogy original drawn from ~\cite{vallor2022ai}}}
Outside of explicit interactions, people may still (consciously or unconsciously) interact with LLM outputs. Even when users are aware they are interacting with these outputs, they may still be subject to the cognitive effects of seamless design. More generally, LLMs and other natural language products are increasingly ubiquitous, and thus increasingly less likely to be noticed. Thus, the synergy between seamless design and ubiquity of LLMs create a unique threat of unmitigated representational harms. 
Seamless communication via natural language bypasses the conscious translation of querying a system. Through ubiquitous deployment, increases in popularity and exposure to LLM outputs can make people more likely to anthropomorphize~\cite{jacobs2023brief}. This, in turn, can miscalibrate user expectations for appropriate functionality, impair their critical reasoning skills, promote misinformation, and increase social disconnection. 
Furthermore, the presentation of a singular, disembodied and seemingly unbiased truth---a distorted reflection of reality---evades recognition and critical evaluation. The promotion of this ``truth'' reflected across unbounded applications entrenches institutional, societal, and social inequalities, as they are amplified through erasure of non-majority truths, representation, and agency.
The ubiquity of LLM deployment facilitates a pervasive, inescapable, and, perhaps, undeniable form of representation. Undesirable representations, stereotypes, and generalizations alone may not be morally reprehensible. Without the proper contextualization, attribution, embodiment, or critical evaluation, however, through internalization or pervasiveness they threaten to become reality.

\section{Proposed Mitigations and Limitations}
\label{Sec::Mitigations}
In the following section, we discuss current solutions and their limitations, followed by some proposed mitigations. We conclude by considering cases in which representational harms may be morally permissible.

\subsection{Current Solutions}
Proposed mitigations involving proportional~\cite{on2002proportionate} or total removal~\cite{Pykes_2023, Schaul_2023, Simonite_2021} of ``harmful'' representations do not properly address nor mitigate the broad spectrum of representational harms. 
First, censorship or the removal of specific content has been shown to be highly ineffective: concealing false positives and inconsistent decisions can lead to increased radicalization, and the shifting of conversations to more lenient platforms~\cite{ottmancensorship}. 
Second, these solutions imply that effects of these associations are zero-sum, and so reducing or eliminating the propagation of negative stereotypes is sufficient to neutralize the representational harm. However, most mental representations are significantly more complex than presupposed by this approach. For instance, the presence of a female criminal in a story (rather than a male) does not thereby eliminate the association of a majority criminals as male, nor does it ensure that the representation of female criminals is realistic (i.e. grounded in social aspects of how each gender is socialized, constructed, and enforced). Thus, the direct replacement of female and male representations cannot simply be ``rebalanced'' in such a way.\footnote{This example also neglects genderqueer and transgender individuals that may or may not lie along the spectrum between male and female.} In addition, the zero-sum approach fails to address the psychological harms of even ``positive'' stereotypes. For example, model minority members may experience increased stress, low self-esteem, invalidation of their feelings, and dissociate with particular facets of their identity due to stereotype threat~\cite{owen2015model, wong2006model, spears1997self}. 

\subsection{Proposed Solutions}

\paragraph{Seamful Design}
One may first consider increasing user agency by means of seamful or frictionful design---that is, deliberate introduction of elements that induce mindfulness, reflection, and critical reasoning skills in users. By increasing user mindfulness and critical engagement with technology, we aim to ensure that users first form opinions of how the technology should behave, what services it should provide, and what principles or values it must adhere to. Although this design approach will not prevent representational harms altogether, it can nonetheless provide first steps towards constructing a system of accountability and public favor for products that mitigate representational harms. Widespread user consciousness about these issues may mean that people will no longer accept the functionality as provided to them, but also take agency over it to demand a higher standard.

Elements of seamful design, such as nudges or other design elements, can also increase transparency and embodiment. Embodiment can highlight the potential flaws, pitfalls, and biases of available technology and more carefully calibrate trust and predicted capabilities. Transparency may facilitate direct lines of communication and accountability to relevant stakeholders (e.g. product-owners and policy makers). Transparency also hinges on many other systemic contingencies, such as public accountability and power (e.g. when disconnected from power, strategic opacity, and false binaries)~\cite{ananny2018seeing}.

For example, a system that conveys multiple candidate responses for any given output may enable a user to get a better understanding of alternative perspectives and discourse within an issue. Visual reminders of the stochasticity of model outputs, though uncomfortable~\cite{McCormick_2016, alieva2023american}, can provide indicators of trustworthiness or representativeness of outputs. This may result in user-dependent trade-offs in utility (e.g., a user whose goals are served by majority responses may appreciate this, whereas those who are not may view this as oppressive/erasure) and should be explored by future HCI research. In addition, deployed systems may employ elements of non-compliance in which users are prompted to reflect on their own norms of content permissibility, diversity of perspectives they wish to be exposed to, and the harms they may be experiencing. Thus, by disrupting user trust, there may also a possibility of gaining it. 

\paragraph{Counter-Narratives}
A specific version of the ``multiple responses'' approach looks to counter-narratives: ``stories that detail the experiences and perspectives of those who are historically oppressed, excluded, or silenced.''~\cite{bergen2023contemporary} 
Rather than manufacturing diversity (e.g., by flipping male and female names and roles, such as a victim for a robber), counter-narratives provide depth, complexity, and liberation through diversity of representation. They allow readers and writers alike to imagine a reality not limited by the historical or current realities of systemic oppression. For instance, the work of~\cite{jones2021we} presents two parallel stories, one that depicts current systems of power and one that embodies abolitionist theories and practices of Black life. They use storytelling and speculative fiction as a methodology to affirm their own lived experience and counteract erasure of less documented experiences. Counter-narratives have even been demonstrated to cause career benefits for minorities in STEM during high-school education~\cite{potvin2023examining}.

Counter-narratives can empower individuals who lack connection to positive representations by challenging dominant narratives and building critical awareness of societal inequities~\cite{grabe2015counter}. The neutralization of representational harms through removal or total lack of engagement by the harmed individual is ineffective and infeasible. Not only does a lack of engagement exacerbate the harms of censorship by most mainstream media, as it often depicts exacerbated institutional oppression and norms~\cite{o2016bechdel}, but it participates in the passive reproduction of such systems in oppressive totality~\cite{kendi2023antiracist} or as ``the way things have always been.'' Instead, facilitating forms of representation that practice the envisioning and constructing of alternative futures, possibilities, and narratives provides liberation to all. Critical perspectives are needed here to determine who decides/creates these (humans (self-identifying members vs. not, as groups are likely to have discourse/disagreement) vs. models) and how they are propagated by or incorporated into training data (either in specific product applications or foundation models).

\paragraph{Measurement for Improvement, Stopped Deployment, and Systems of Accountability}
One clear implication from our work is that it is imperative to measure these potential representational harms at a variety of granularities. Given such measurements, relevant stakeholders can make an informed decision whether or not to deploy (or continue deploying) a piece of technology. Such decisions can be based on a variety of reasons, including the scale or scope of the representational harms, the effects on public opinion/favor, or the effects on public good. Given the methods laid out in this paper (see e.g. \cref{Sec::Measurement}), it is no longer sufficient to claim ignorance. Measurement frameworks must be deployed so that the appropriate mitigations (including lack thereof) can be employed. Institutional and organizational processes can be implemented to ensure systems of accountability and appropriate reparations. Thus, for instance, harm measurement may lead to pipeline impacts such as retraining, improved screening of outputs, and blocks on certain kinds of queries.

\subsection{Morally Defensible Cases}
Interestingly, and perhaps counter-intuitively, there may be some cases in which representational harms are morally permissible. Most simply, those representational harms may be outweighed (morally) by other considerations (though we emphasize that we are not endorsing simple cost-benefit analyses). For example, suppose a company's LLM has a tendency to impose incorrect pronouns on users~\cite{ovalle2023m}. While this behavior may certainly have the potential to cause more harm to some than others (through misgendering people who are already often misgendered, such as members of the trans community), the company may have certain freedoms of expression that would be harmed if they were forced to change the LLM outputs. That is, we need to consider the possibility that attempts to mitigate representational harms may create other harms (or risks) elsewhere. 

More interestingly, representational harms may be the ``cost'' that must be borne to achieve a more ethical (global) state. In general, any ethical analysis to guide mitigations must consider harms relative to the moral baseline, rather than the status quo. If the status quo is morally problematic, then we may need to change people's representations, but those changes will often constitute ``harms'' for some groups. Of course, those latter harms may be morally justified as a permissible (perhaps only) way to achieve a more ethical (overall) state. Nonetheless, we should not overlook the possibility that this process of change may yield representational harms for some individuals or groups as an unavoidable price. 

This balancing process is perhaps easiest to see through an example. A query to an LLM about what a ``Computer Scientist'' looks like might output all White or Asian males, as they are currently the majority in the field. This might cause representational harms towards women or non-binary individuals. When appropriate, a mitigation might alter the gender composition of the output, causing male individuals to not receive the same representational benefits as they might have with a homogeneous output. More heterogeneous outputs would potentially lead males to be less likely to think of themselves as potential computer scientists. However, in this case, one must consider the extent to which users are \emph{justly} entitled to their current representations (and corresponding potential benefits). Male users are arguably not legitimately entitled to the representational (and allocative) benefits that they receive by virtue of the homogeneous outputs from the LLM, as those benefits arise merely from (biased) historical practices, rather than morally relevant attributes. And of course, the illegitimacy of the benefits is exacerbated by the empirical inaccuracy of a homogeneous representation (e.g., Ada Lovelace was the first computer programmer; women, non-binary, and trans people work within the computer science field today; and so on). Therefore, men are receiving a representational benefit above that of the moral baseline (which should presumably encode that all people with certain skills and knowledge can be computer scientists), it may be morally permissible to remove this unwarranted benefit. 

We acknowledge that removal of such representational benefits may be quite complex. In many of these cases, representational harms operate through the vehicle of public presentation, and so we must be sensitive to empirical details about how those presentations are consumed. For example, presentation of 100 male computer scientists followed by 100 female scientists and 100 genderqueer computer scientists is unlikely to lead to balanced representations since most people will only see the first few. More generally, there may sometimes be no way to reduce potential representational harms to one individual or group without imposing some harms on a different group. In such cases, we need to think carefully about the relevant moral baseline, as moves towards that baseline may be morally justified and permissible (though we do not suggest that such moves are always obligatory).

\section{Conclusion}
Many authors have noted the distinction between allocative and representational harms, only to then focus the bulk of their attention on the former. In this paper, we have focused on core foundational issues about representational harms: how are they defined, what is left out, how are they measured, how can they be mitigated, and why all of these matter. Our work constructs a framework for characterizing representational harms. We have provided some key types of such harms, but we emphasize that our tables are certainly not comprehensive. Nonetheless, we contend that our framework provides scaffolding to support research that expands the types of representational harms, including associated measures and mitigations. In addition, we provide high-level formulations for practitioners to understand the expertise and future research required to implement measurement practices that can effectively identify and track representational harms. Through a case study and identification of LLM characteristics, we motivate why these measures are paramount to properly assess and justify deployment of public-facing algorithmic outputs. Finally, we detail further considerations when representational harms may be inevitable and morally permissible. One key insight from our examination of potential mitigation strategies is that we are not limited to merely diversifying outputs of algorithms, but instead can find more innovative mitigations that move everyone closer to the relevant moral baselines.

\newpage
\printbibliography

\section{Tables of Non-Behaviorist Representational Harms}

Below we provide some additional examples of changes in affective and cognitive states and examples of measures that may be correlated with representational harms (see e.g. \cref{Table::Individual Non-Behavioral Rep Harms} and \cref{Table::Group Non-Behavioral Rep Harms}).

\begin{table}[htbp]
    \centering
    \begin{tabular}{p{2cm}|p{8cm}|p{3cm}}
        \toprule
        \textbf{Theme} & \textbf{Phenomena} & \textbf{Measure of Phenomena}\\
        \midrule
        Possibilities, Option Consideration (Individual)
        & \begin{enumerate}
            \item  Restriction of diversity of roles / actions / possibilities (reduce the scope of options one might consider)
            \item Compartmentalization of identity attributes (and associated characteristics) and where they are declared/shared
            \item Denial of discontent/apathy/depression due to narrow options 
            \item Increased feelings of hopelessness, defeat, apathy, or nihilism
        \end{enumerate}
        & \begin{enumerate}
            \item \cite{akhtar2020covid}
            \cite{lent2002social, lent2006conceptualizing}
            \cite{craig1986different, adams1979toward}
            \item 
             
            \cite{phinney1990ethnic, phinney1992multigroup, singh1977some, crawford2002influence, craig1986different, adams1979toward}
            \item \cite{clark1999racism, fernando1984racism, seligman1975helplessness}
            \item \cite{peterson1993learned, clark1999racism, seligman1975helplessness}
        \end{enumerate}\\
        \midrule
        Stress (Individual) 
        & \begin{enumerate}
            \item Changes in perceived danger/threat/risk 
            \item Perceptions of control
            \item Hyper-awareness of the self -- increased perception of threats/stress/anxiety/guilt/disappointment to implicit/explicit stereotype defiance and compliance
            \item Increased pressure/stress to conform to a stereotype or perceived lack of conformity with the stereotype
            \item Extensive/over-analysis of whether specific actions will be perceived (pos/neg), sometimes accompanied by only enacting these actions when a positive result is predicted
            \item Increased perceived pressure to ``succeed'' under traditional norms for success (financial, recognition) as opposed to abstract ones like happiness/content 
            \item Distorted perception of the self -- increased predisposition to negative self talk
            \item Strategic or otherwise decreased memory of traumatic or high stress periods
        \end{enumerate}
        & \begin{enumerate}
            \item \cite{wolff2019define}
            \item \cite{peterson1993learned, clark1999racism, seligman1975helplessness}
            \item \cite{akhtar2020covid, bandura1985catecholamine} 
            \item 
            \item 
            \item \cite{clark1999racism}
            \item 
            \item 
        \end{enumerate}\\
        \midrule
        Social (Individual)
        & \begin{enumerate}
            \item Increased likelihood to personalize/attribute negative interactions to identity characteristics
            \item Perception of injustice
            \item Predisposition to initiate interactions with people of similar or marginalized identity characteristics, assumption of shared experiences and solidarity
            \item Anxiety and hesitation around speaking up, contributing to a conversation
            \item Distorted perception of the self -- minimization of achievements, imposter syndrome
            \item Decreased feelings of self-importance and value, particularly wrt contributions within social settings
            \item Increased dependence on validation from positions of authority for ``non-traditional'' paths
            \item Rejection sensitivity
            \item Hyper-independence -- increased likelihood of do-it-yourself attitudes to the point of self-detriment towards things that may be otherwise achieved faster/more efficient to ask for help
            \item Narrowed definitions of self-worth and value based on contribution/devotion/service to others
            \item Contempt, undeserved hate/projection (or increased predisposition) towards stereotype non-conforming/defying individuals
            \item Disproportionate rewards / positive valence emotions for conforming to stereotypes
        \end{enumerate}
        & \begin{enumerate}
            \item \cite{clark1999racism}
            \item \cite{neumann2021development}
            \item \cite{phinney1990ethnic}
            \item 
            \cite{larkin1998cardiovascular, pearlin1989sociological}
            \item 
            \item \cite{clark1999racism}
            \item 
            \item 
            \item \cite{akhtar2020covid, clark1999racism}
            \item \cite{clark1999racism}
            \item \cite{clark1999racism}
            \item 
        \end{enumerate}\\
        \bottomrule
    \end{tabular}
    \caption{Definitions of Representational Harms and Associated Applications}
    \label{Table::Individual Non-Behavioral Rep Harms}
\end{table}

\begin{table}[htbp]
    \centering
    \begin{tabular}{p{2cm}|p{8cm}|p{3cm}}
        \toprule
        \textbf{Theme} & \textbf{Definition} & \textbf{Metric}\\
        \midrule
        (Within Group)
        & \begin{enumerate}
            
            \item Increased sensitivity, awareness, and/or judgement of other group-members rejection/conforming to stereotypes (i.e. familial expectations)
            \item Decreased sense of connection, responsibility, and solidarity within a group or with a group identity
            \item Detachment, ambivalence, or resentment towards within-group causes
            \item Greater predisposition to fragmentation or fragility within group connections
            \item Increased likelihood of intolerance of discourse/dissenting opinion, narrowed/prioritized (+potentially polarized) subgroups within a group
            \item Increased tensions, resentment, ostracization from a community for ``bad reputation''/high risk individuals
            \item Increased likelihood of acceptance of non-identifying members into the community for their reputation rehabilitation potential, this can also cause downstream harms as it may be perceived as ``pushing out'' similar/equivalent members (i.e. my caretaker accepts my different race partner into the family because of their career succcess, but treats me poorly because of my lack of traditional career success -- both are seemingly equivalent on the basis of age)
        \end{enumerate}
        & \begin{enumerate}
            \item 
            \item 
            \cite{phinney1990ethnic, singh1977some}
        
            \cite{craig1986different, bullock1987perceptions}
            \item 
            \cite{phinney1990ethnic, singh1977some, bullock1987perceptions}
            \item 
            \item \cite{singh1977some}
            \item 
            \item 
        \end{enumerate}\\
        \midrule
        (Between Groups)
        & \begin{enumerate}
            \item Increased animosity, resentment, feelings of competition between (marginalized) groups
            \item Greater disconnectivity, lack of support between groups
            \item Increased feelings/perception of singularity in success -- ``there can be one successful [insert group category]'' -- this can be among groups, within a group, between individuals
            \item Predisposition/implicit reinforcement of perceived hierarchies in financial, career success and/or happiness
            \item Decreased likelihood or hope for collaborations and intergroup support/collaborations
        \end{enumerate}
        & \begin{enumerate}
            \item \cite{berry1986assessment}
            \item 
            \cite{singh1977some, berry1986assessment, bullock1987perceptions}
            \item 
            \item 
            \item
            \cite{singh1977some, berry1986assessment}
        \end{enumerate}\\
        \midrule
        (Indirect?)
        & \begin{enumerate}
            \item Feeling calm/relaxed in otherwise stressful situations due to historically comparably ``worse''/more stressful situations
        \end{enumerate}
        & \begin{enumerate}
            \item 
        \end{enumerate}\\
        \bottomrule
    \end{tabular}
    \caption{Definitions of Representational Harms and Associated Applications}
    \label{Table::Group Non-Behavioral Rep Harms}
\end{table}

\end{document}